\documentclass{article}
\usepackage{graphicx} % Required for inserting images
\newcommand\blfootnote[1]{%
  \begingroup
  \renewcommand\thefootnote{}\footnote{#1}%
  \addtocounter{footnote}{-1}%
  \endgroup
}
\usepackage[a4paper, total={7in, 9in}]{geometry}
\usepackage[
backend=biber,
style=chem-angew,
]{biblatex}
\usepackage{placeins}

\title{Elucidating nanostructural organisation and photonic properties of butterfly wing scales using hyperspectral microscopy}

\author{Anna-Lee Jessop$^{1,\dagger}$, Primo\v{z} Pirih$^{2,\dagger}$, Limin Wang$^{2}$, \\ Nipam H. Patel$^{3,4}$, Peta~L.~Clode$^{5,6}$, \\ Gerd E.~Schr\"oder-Turk$^{1,7,*}$, and Bodo~D.~Wilts$^{2,*}$}

\begin{document}

\maketitle

\section{Abstract}
Biophotonic nanostructures in butterfly wing scales remain fascinating examples of biological functional materials, with intriguing open questions in regards to formation and evolutionary function. One particularly interesting butterfly species, \textit{Erora opisena} (Lycaenidae: Theclinae), develops wing scales that contain three-dimensional photonic crystals that closely resemble a single gyroid geometry. Unlike most other gyroid forming butterflies, \textit{E. opisena} develops discrete gyroid crystallites with a pronounced size gradient hinting at a developmental sequence frozen in time. Here, we use a hyperspectral (wavelength-resolved) microscopy technique to investigate the ultrastructural organisation of these gyroid crystallites in dry, adult wing scales. We show that reflectance corresponds to crystallite size, where larger crystallites reflect green wavelengths more intensely; this relationship could be used to infer size from the optical signal. We further successfully resolve the red-shifted reflectance signal from wing scales immersed in refractive index oils with varying refractive index, including values similar to water or cytosol. Such photonic crystals with lower refractive index contrast may be similar to the hypothesized nanostructural forms in the developing butterfly scales. The ability to resolve these fainter signals hints at the potential of this facile light microscopy method for \textit{in vivo} analysis of nanostructure formation in developing butterflies.

\blfootnote{$^1$School of Mathematics, Statistics, Chemistry and Physics, Murdoch University, Australia.} 
\blfootnote{$^2$Chemistry and Physics of Materials, University of Salzburg, Austria.} \blfootnote{$^3$Marine Biological Laboratory, University of Chicago, USA.} 
\blfootnote{$^4$Centre for Microscopy, Characterisation, and Analysis, University of Western Australia, Australia.} 
\blfootnote{$^5$School of Biological Sciences, University of Western Australia, Australia.} 
\blfootnote{$^6$Research School of Physics, The Australian National University, Australia.}
\blfootnote{$\dagger$ Contributed equally.}
\blfootnote{*Authors for correspondence: G.E. Schr\"oder-Turk: g.schroeder-turk@murdoch.edu.au; B.D. Wilts: bodo.wilts@plus.ac.at }

\newpage

\section{Introduction}
\begin{figure}[h] 
    \centering
    \includegraphics[width=\textwidth]{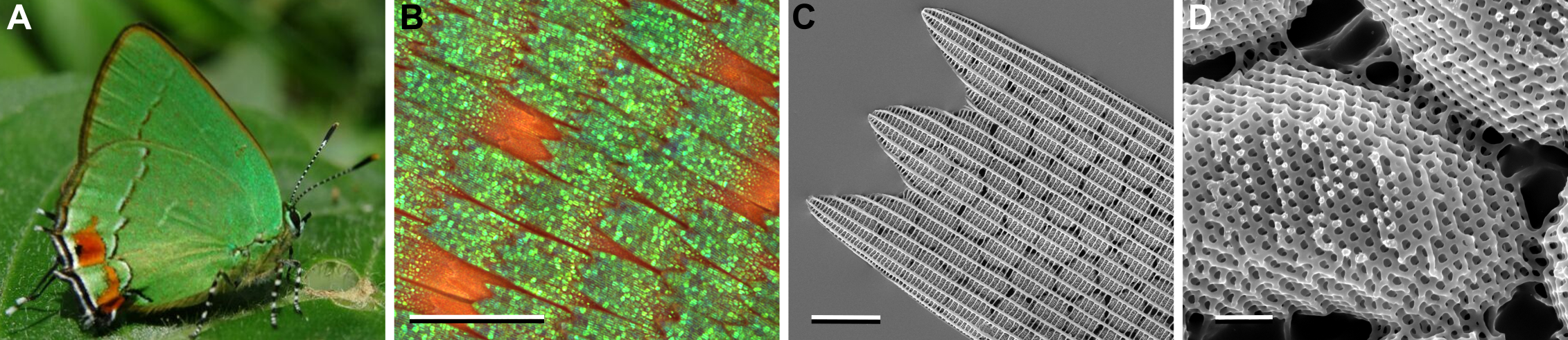}
    \caption{The green wings of \textit{E. opisena} originate from wing scales containing photonic gyroid nanostructures. (A) Photograph of \textit{E. opisena}, reproduced with permission from P. Brodkin. (B) Light microscopy image of the wings of \textit{E. opisena}. (C) Scanning electron microscopy image of a single green scale showing the ridges, crossribs, and discrete gyroid crystallites within the scale lumen beneath. (D) Gyroid crystallites imaged from the underside of the scale with the lower lamina removed. Scale bars: B) 100 $\mu$m, C) 10 $\mu$m, D) 1 $\mu$m.}
    \label{fig:Figure1}
\end{figure}

Some of the most visually stunning displays in nature are created by structural colours. This phenomenon arises as a result of the interference of light with periodic nanostructures and is known to occur across almost all kingdoms of life including animals \cite{wilts2012brilliant, teyssier2015photonic, lythgoe1989structural, thayer2023meta}, plants \cite{vignolini2012pointillist, chandler2015structural, van2014iridescent}, bacteria \cite{schertel2020complex}, and protists \cite{dolinko2012photonic,bauernfeind2024thin} (for reviews, see \cite{vukusic2003photonic, srinivasarao1999nano, kinoshita2008structural, parker2000515}). Among those, butterfly wings are arguably the most well recognized example of structural colour in nature. Butterflies employ a diverse set of nanostructural morphologies within their wing scales that behave as photonic structures \cite{thayer2023meta,kinoshita2008structural}. The single gyroid crystals found in numerous green butterflies \cite{michielsen2008gyroid,Saranathan2010,schroder2011chiral,WiltsScienceAdv2017,yoshioka2014polarization,wilts2012iridescence} are one of the more intriguing nanostructural morphologies, comprising a highly ordered, chiral and sponge-like network with cubic crystal symmetry \cite{hyde2008short} (Figure~\ref{fig:Figure1}). In butterflies, these gyroid photonic crystals reflect green light that likely functions in camouflage, as an aposematic signal, and in mate choice. Biophotonic single gyroid structures have also been reported in weevils \cite{Saranathan2015Nanolet} and birds \cite{Saranathan2021} with slightly shorter and longer reflectance wavelengths compared to the butterflies, respectively.

The development of butterfly nanostructures occurs during pupation. Each wing scale develops from a single epithelial cell that secretes a chitinous cuticle that follows the template of the cell's plasma membrane \cite{overton1966microtubules, greenstein1972ultrastructure}. Based on Ghiradella's seminal ultramiscroscopic studies \cite{ghiradella1974development,ghiradella1989structure,GHIRADELLA2010135}, the prevailing hypothesis of gyroid formation in butterflies is that the wing scale cell's smooth endoplasmic reticulum adopts a gyroid morphology and acts as a template to the nascent chitin polymer \cite{Saranathan2010, ghiradella1989structure,WiltsScienceAdv2017,Winter2015PNAS,wilts2019nature}, while the whole scale cell is still living and containing cytosol. In parts due to the inability to observe nanostructure growth in butterflies {\it in vivo} several aspects of this formation or growth process remain open or speculative.

The optical properties of single gyroid photonic crystals are well understood, through band structure analyses \cite{SabaCrystalsCorrectBandStructure2015,MoldovanThomas2002,MichielsenKole2003,saba2014absence} and numerical reflectance calculations using finite-difference time-domain methods \cite{MichielsenRaedtStavenga2010,SabaCrystalsCorrectBandStructure2015,saba2014absence}, plane-wave expansions \cite{BabinHolyst2002,MoldovanThomas2002,Saranathan2010}, or scattering matrix methods \cite{SabaPRL2011}. Single gyroid optical properties have also been analysed through experiments on custom fabricated 3D gyroid geometries on the centimeter scale \cite{pouya_electromagnetic_2012,PouyaVukusicTunableGyroid} or nano/micrometer scale \cite{gan2016biomimetic,PengMicron3DPrintGyroidReflectance2016,turner2013miniature}, see \cite{dolan2015optical} for a review. The optical observations in gyroid biophotonic crystals in butterflies \cite{Saranathan2010, saba2014absence, singer2016domain,yoshioka2014polarization,wilts2012iridescence} or replicas thereof \cite{mille_3d_2013} are largely explained as photonic crystal effects, with or without pigmentary absorption. It is well known that the immersion of dielectric photonic crystals in media with refractive indices different to that of air (\textit{n}=1) changes the color of their reflectance, used in sensor applications \cite{kertesz2018optical,potyrailo2013discovery, potyrailo2007morpho}; the changes in coloration of butterfly wings immersed in alcohol or other liquids is a popular demonstration experiment \cite{BoberButterflyEthanolChemEd2018}. If indeed a chitin gyroid structure was present in the pupae during development, it would be immersed within the cytosol ($1.36 < n < 1.39$, \cite{liu2016cell}) leading to a change in reflected color compared to the dry state. Indeed, the transition of some green butterflies from a bronze color state just after emerging to a bright green coloration is likely to be due to this effect \cite{DryOutButterflyAber2024}.

The focus of this study is on the green wing scales of the Neotropical lycaenid butterfly, \textit{Erora opisena} (formerly \textit{Thecla opisena}; Figure~\ref{fig:Figure1}). These wing scales contain single gyroid nanostructures that, remarkably, form disjoint crystallites with a pronounced size gradient along the scale \cite{WiltsScienceAdv2017} (a similar crystallite structure was observed in {\it Mitoura\,grynea} \cite{ghiradella1989structure}). Gyroid crystallites towards the tip of the scale are larger and gradually decrease in size towards the base of the scale. Wilts {\it et al.} \cite{WiltsScienceAdv2017} hypothesised that this size gradient may be a developmental sequence frozen in time and that the formation of the crystallites could be through a growth or extrusion process. If this were the case, one would expect the gyroid photonic crystals to begin developing initially as small crystallites towards the tip of the wing scale, growing larger over time. In this sense, the optical properties of the smaller gyroid crystallites found towards the base of the adult wing scales may be analogous to those of crystallites at an early stage of development.

Here, we use hyperspectral microscopy (HSM), and auxiliary scanning electron microscopy, X-ray tomography, and optical modelling, to explore the relationship between gyroid crystallite size and reflectance and the effect of immersing wing scales in refractive index-matching oils with refractive indices similar to cytosol. Hyperspectral microscopy is a microscopy technique that provides spectral data for each image pixel \cite{Schultz2001}. This differs from conventional light microscopy that only records in three colour channels (red, blue, and green) and from multispectral imaging that records in three to fifteen colour channels. Hyperspectral microscopy, on the other hand, can record in many hundreds of colour channels, depending on the HSM set up (here, we record from 30 channels). Hyperspectral microscopy also differs from the more conventional microspectrophotometry (MSP) that also records a full spectrum but can only record from a single location at a time, whereas HSM allows spectra to be recorded from many locations simultaneously. This advantage enables efficient reflectance measurements to be collected across a large area, such as an entire wing scale. As such, HSM is a method that combines the spatial resolution of light microscopy with high spectral resolution, as an alternative to MSP.

Hyperspectral microscopy, and hyperspectral imaging more broadly, have been applied across many different fields (see the reviews in \cite{banu2023hyperspectral, roth2015hyperspectral}) providing a method for measuring spectra across small spatial scales. To date, HSM has already been used to elucidate colour generation in adult butterfly wings \cite{MedinaVukusic2011}, for species discrimination in \textit{Drosophila} \cite{takahashi2023quantitative}, coating quality control \cite{medina2011scattering}, \textit{in vivo} measurements of human irises \cite{medina2011hyperspectral}, for angular-dependent reflectance of iridescent butterflies \cite{medina2014detailed}, for determination of screening pigments in the eye \cite{blake2019compound,ilic2022simple}, and to measure defects in self-assembled cellulose nanocrystals \cite{ZhuVignolini2020}. 

While HSM, like all conventional light microscopy methods, cannot resolve the biophotonic nanostructures, HSM can resolve the optical reflections the structure creates. This article demonstrates that this method is sufficiently sensitive and accurate to resolve finer ultrastructural details of the photonic signal of butterfly wing scales. This includes resolving signals in situations (namely in immersion experiments) that mimic the situation one may find in a developing wing scale, hence pointing at the potential of HSM for future applications in imaging biophotonic nanostructural development \textit{in vivo}.

\section{Results}
\begin{figure}[h]
    \centering
    \includegraphics[width=14cm]{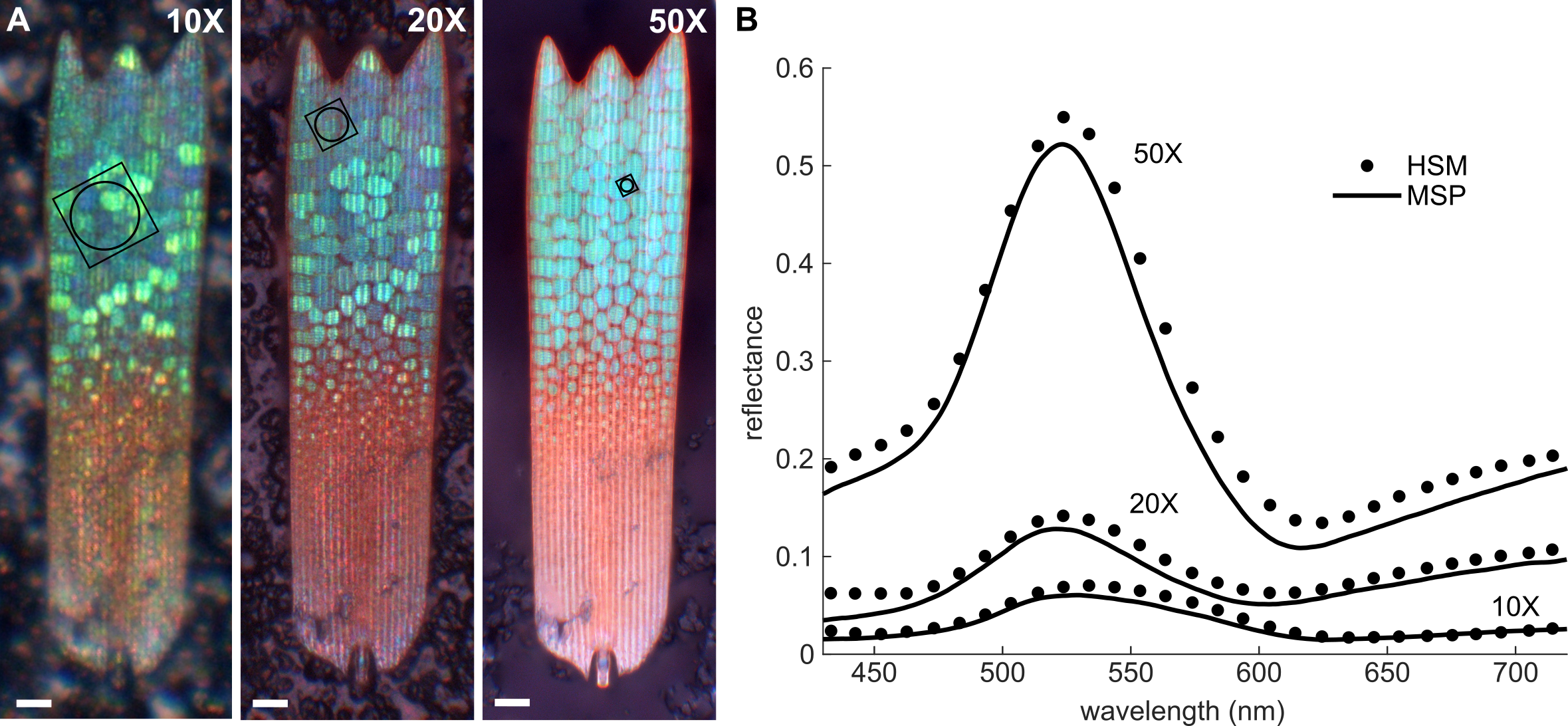}
    \caption{Reflectance spectra from small patches of a single green wing scale from \textit{E. opisena} measured using hyperspectral microscopy (HSM) and microspectrophotometry (MSP). (A) Light microscopy images of a single wing scale taken with 10$\times$ (NA=0.30), 20$\times$ (0.60) and 50$\times$ (0.95) objectives, showing locations of measurement areas producing the results shown in (B). The colour of crystallites changes with the numerical aperture of objectives. (B) Reflectance spectra from each location in (A) measured using HSM (dots) and MSP (lines). The scale bars in (A) are 10 $\mu m$, the circles' sizes represent the measurement area for MSP and the square sizes represent the measurement area for HSM (10000 camera pixels).}
    \label{fig:Figure2}
\end{figure}

\textit{Erora opisena} is a small Lycaenid butterfly with an overall size of $\approx$2\,cm, native to the Northern Neotropics \cite{WiltsScienceAdv2017}. The ventral sides of both wings of the butterfly are vivid green, with the hindwings containing small red and white patches (Figure~\ref{fig:Figure1}A). The green colour arises from wing scales that contain single gyroid photonic nanostructures within the scale lumen (see Figure 1 of \cite{WiltsScienceAdv2017} and Figure~\ref{fig:Figure1}B--D). Single scales show a diminishing gradient of green colouration from the scale tip to middle, whereas the lower half is red-coloured and no photonic nanostructure is present (Figure~\ref{fig:Figure2}A). When attached to the wing, the wing scales partially overlap and the reddish parts are mostly hidden (Figure~\ref{fig:Figure1}B). 

Observing the scale through microscope objectives with increasing numerical apertures (10$\times$, NA=0.30; 20$\times$, NA=0.60; 50$\times$, NA=0.95), the green colour of crystallites changes in hue, desaturates, and increases in intensity and uniformity (Figure~\ref{fig:Figure2}B). A similar NA-dependent effect, which is due to illuminating the sample and collecting the light from ever larger spatial angles, has been previously observed in the gyroid structures of the butterfly, \textit{Callophrys rubi} \cite{corkery2017colour}. Using HSM and MSP, we measure reflectance (spectra) as a measure of the relative reflectance to a mirror standard, see Methods section.

We used the green, red, and white scales of \textit{E. opisena} to compare the spectra obtained with HSM and MSP methods (see Figure~\ref{fig:Figure2}B and Figure S1). HSM was performed between 420--720 nm with a wavelength resolution of 10 nm (see Methods), while MSP was performed between 380--760 nm. To directly compare the two methods, the same locations and approximately equal measurement areas were used (see Figure~\ref{fig:Figure2}A). The MSP measurement areas were $\approx$ 320, 78, 14 $\mu m^2$ for the 10$\times$, 20$\times$, 50$\times$ objectives, respectively. The corresponding HSM square measurement area (100$\times$100 camera pixels) could in principle be further reduced to 10$\times$10 or less pixels, achieving  diffraction-limited measurement area <1 $\mu m$ with a 50 $\times$ objective. 

The reflectance curves obtained with HSM (dots) and MSP (lines) were closely matching for green, red, and white \textit{E. opisena} wing scales (Figure \ref{fig:Figure2}B; Figure S1). The peak wavelength for green scales measured by HSM with 10$\times$, 20$\times$, and 50$\times$ objectives occurred at 533 nm, 523 nm, and 523 nm, respectively, differing by less than 1\% from MSP measurements (528 nm, 521 nm, and 523 nm, respectively). Reflectance amplitudes also agreed well, differing on average over wavelengths by 16\%, 20\%, and 23\%, respectively. For the red and white scales, the MSP and HSM reflectance measurements differed by 4\% and 1\%, respectively (Figure S1). 

\subsection{Crystallite size determines reflectance amplitude}

\begin{figure}[h!]
    \centering
    \includegraphics[width=14cm]{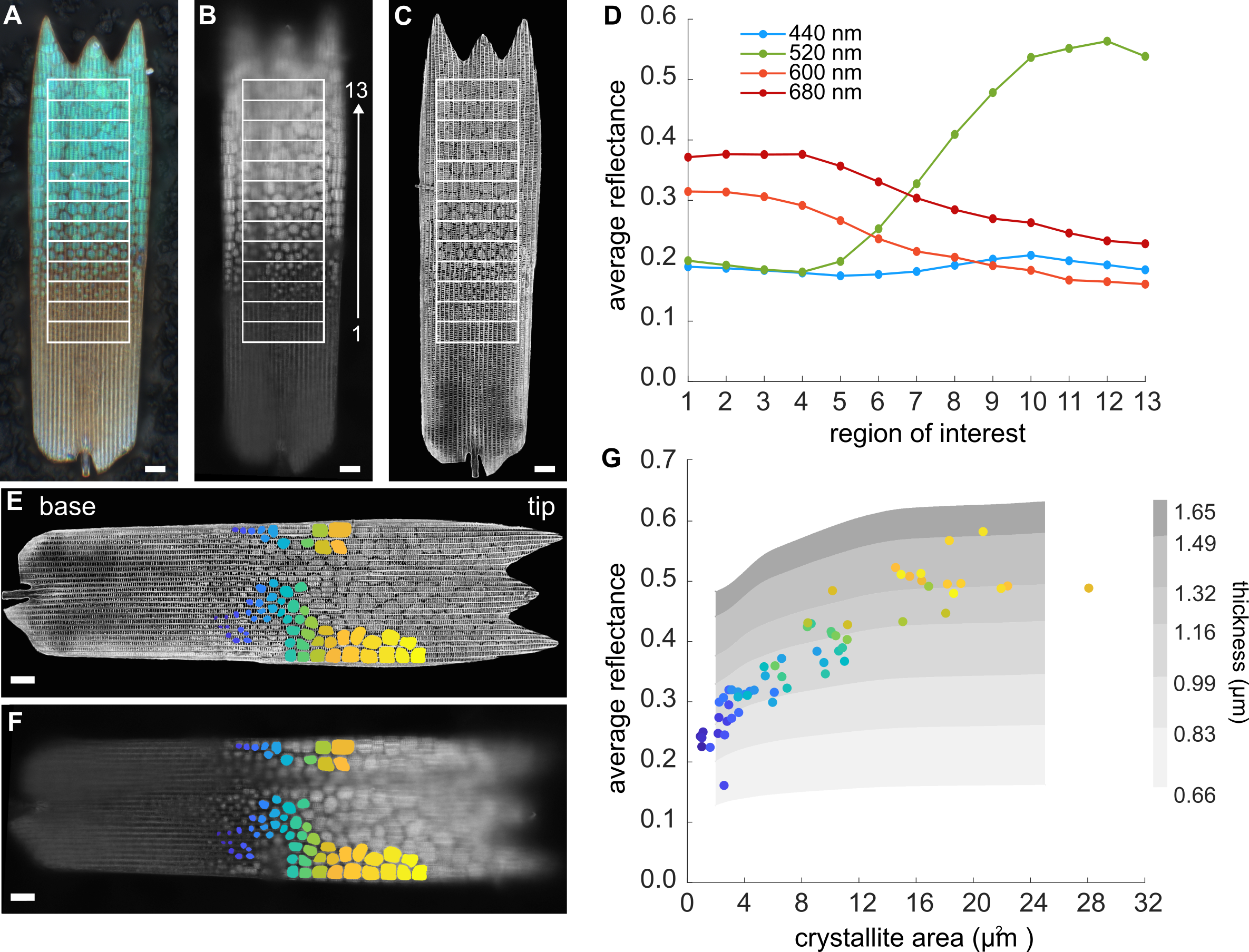}
    \caption{Reflectance increases with crystallite size across a single wing scale. (A--C) Microscopy images of a single green \textit{E. opisena} wing scale showing the locations of 13 regions of interest that were used to measure relative reflectance from the base to the tip of the scale. (A) Colour image taken under white light illumination. (B) Monochromatic image at 520 nm. (C) Scanning electron microscopy (SEM) on-view image of the same scale. The crystallites are visible below the array of ridges and cross-ribs. (D) Average reflectance across 13 regions of interest measured using hyperspectral microscopy (HSM). Coloured points show the average reflectance under 440 nm (blue), 520 nm (green), 600 nm (orange), and 680 nm (red) light. (E,F) The locations of segmented crystallites from a single scale superposed on the SEM image (E) and the monochromatic light microscopy image (under 520 nm illumination). (G) The average reflectance of 520 nm light across each segmented crystallite versus the area of each crystallite. The colours of each point correspond to the colours of each segmented crystallite in panel E and F. Grey shading shows the theoretical reflectance obtained from FDTD simulations of single gyroid geometries with a refractive index equal to 1.56, a lattice parameter of 330 nm, a solid volume fraction of 0.3, areas between 1.96 $\mu$m$^2$ and 25 $\mu$m$^2$ and thicknesses between 0.66 $\mu$m and 1.65 $\mu$m.}
    \label{fig:Figure3}
\end{figure}

The lumen of \textit{E. opisena} green scales contains discrete gyroid crystallites that exhibit a pronounced size gradient with smaller crystallites towards the middle and larger crystallites at the tip \cite{WiltsScienceAdv2017} (Figure~\ref{fig:Figure3}A--F). HSM enabled us to efficiently investigate whether there is a relationship between crystallite size and reflectance. For this, we measured the average reflectance across 13 regions of interest (200 $\times$ 800 pixels) beginning from the base of the scale (where no photonic nanostructure is present) and ending at the tip of the scale (where the largest crystallites are located) (Figure~\ref{fig:Figure3}A--C). The reflectance at 520 nm increased from the base ($\approx 0.20$) to the tip ($\approx 0.55$) of the scale. The opposite trend was observed at 600 and 680 nm, where the reflectance was higher towards the base of the scale. Reflectance at 440 nm remained approximately constant across the scale length. 

Segmentation of individual crystallites was based on the SEM image (Figure~\ref{fig:Figure3}E). This allowed identification of the exact locations of the structures without the uncertainty caused by the optical effects of the structures. We further analysed only the crystallites that appeared sharp in the HSM image stack. In total, 50 crystallites were segmented, ranging in size from 0.9 $\mu$m$^2$ to 28 $\mu$m$^2$, with smaller crystallites displayed in blue and larger crystallites displayed in yellow (Figure~\ref{fig:Figure3}E--F). The mean reflectance at 520 nm was calculated for each crystallite (dots in Figure~\ref{fig:Figure3}G, the colours consistent with Figure~\ref{fig:Figure3}E--F). The mean reflectance of crystallites increased with their size and ranged from $\approx$0.16 for the smaller crystallites to $\approx$0.58 for the larger crystallites (Figure~\ref{fig:Figure3}G).

The measured trend is supported by optical modelling of idealised chitin gyroid structures. We modelled the reflectance of idealised chitin gyroid structures with thicknesses varying between 0.66 $\mu$m and 1.65 $\mu$m, and areas varying between 1.96 $\mu$m$^2$ and 25 $\mu$m$^2$. A positive correlation was found between reflectance and crystallite area (grey shading in Figure~\ref{fig:Figure3}G), however, increasing only the area does not fully explain the experimental results. The modelling suggests that both crystallite area and thickness determine the absolute reflectance of the crystallites. This hints that in the green scales of \textit{E. opisena},  the larger crystallites at the tip of the scale could be $\approx$ 50\% thicker than the smallest crystallites towards the scale base, based on the readings 1.3--1.5 $\mu m$ and 0.8--1.1 $\mu m$ for large and small crystallites, respectively. To further investigate this, we measured the thickness and area of gyroid crystallites from an X-ray tomogram of a different \textit{E. opisena} wing scale (Figure S3). Crystallites that covered a larger area were indeed also thicker, with the largest measured crystallites (area$\approx$40 $\mu m^2$) having a thickness of $\approx$2.5 $\mu m$ and the smallest measured crystallites (area$\approx$3 $\mu m^2$) having a thickness of $\approx$0.5 $\mu m$ (Figure S3). These data are consistent with thickness estimates that can be extracted from the reflectance and optical modelling data presented in Figure~\ref{fig:Figure3}G.

\subsection{Reduced refractive index contrast causes dimmed, red-shifted reflectance}

Photonic crystals in mature insect scales are usually composed of a network of dry, chitinous cuticle and a network of air \cite{michielsen2008gyroid,Saranathan2010,wilts2012brilliant,WiltsScienceAdv2017}, resulting in a refractive index contrast between the two networks of $\approx$1.55 \cite{leertouwer2011refractive}. The situation is quite different in the growing scales, where the refractive index contrast is much lower, since one network is cytosol with a refractive index $1.36 < n < 1.39$ \cite{liu2016cell} and the other is likely hydrated chitin, $n < 1.55$. To mimic these conditions, we can replace the air in the network with a fluid of known refractive index. We therefore used HSM to measure the reflectance of isolated wing scales immersed in different refractive index oils.

\begin{figure}[h!]
    \centering
    \includegraphics[width=14cm]{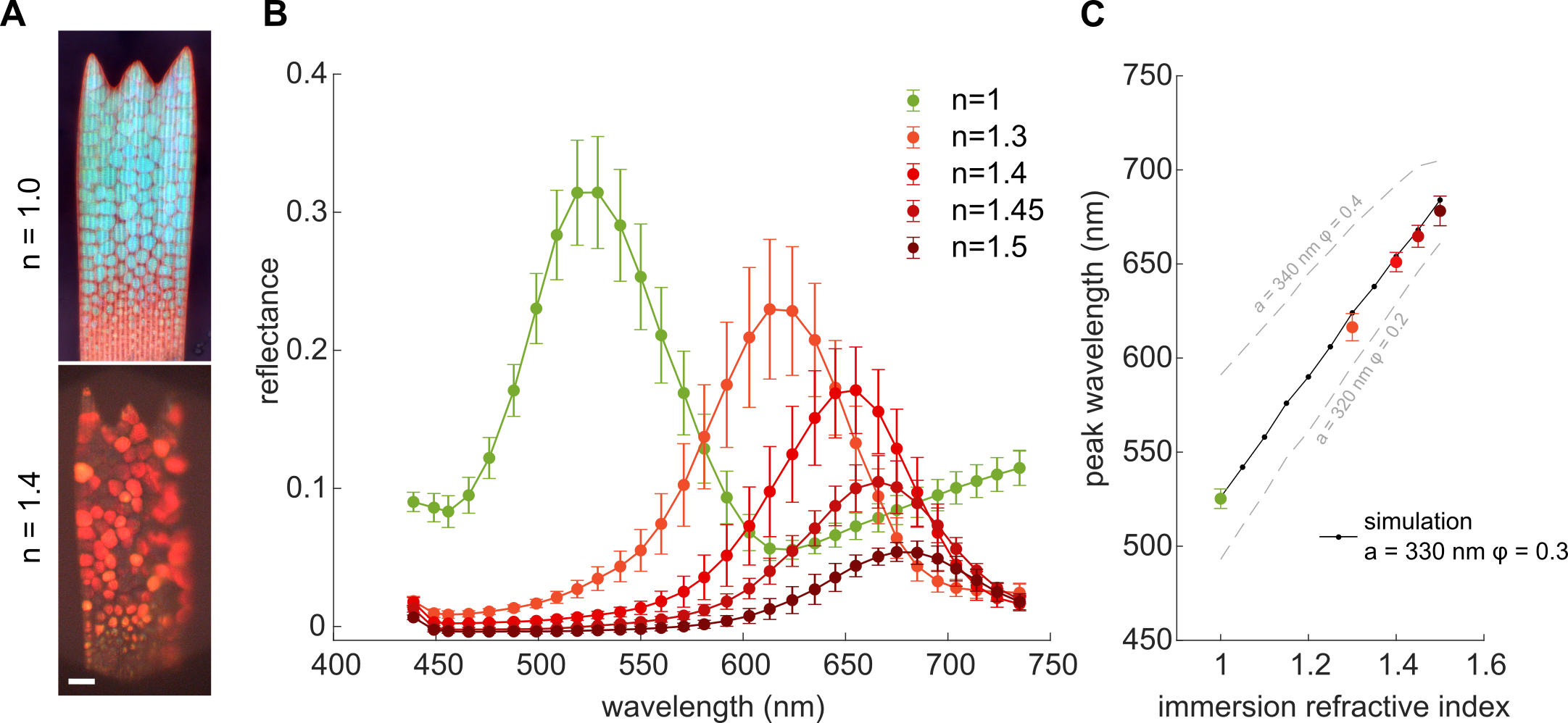}
    \caption{Change in reflectance spectra of \textit{E. opisena} wing scales upon immersion in refractive index matching oils. (A) Light microscopy images of a green wing scale in air (\textit{n} = 1.00, upper panel) and immersed in oil with refractive index \textit{n} = 1.40 (lower panel). (B) Reflectance curves of ten crystallites (mean$\pm$SD) from a single green wing scale in air (green) and from scales immersed in oils with refractive indices of 1.30 (orange), 1.40 (light red), 1.45 (medium red), and 1.50 (dark red). Measurements for each crystallite were averaged from areas containing 1600 camera pixels. (C) The peak reflectance wavelength (mean$\pm$SD) for each immersion experiment (coloured points, mean$\pm$SD) along with the simulated peak reflectance wavelengths determined for theoretical immersion refractive indices of 1 through to 1.50 (black and grey lines). FDTD simulations were conducted on a single gyroid geometry within a range of values for the lattice parameter $a$ and volume fraction $\phi$. The combination of $a=330nm$ and $\phi=0.3$, which provides a close agreement with the measured data, fits with estimates from electron microscopy (black points and line). The combinations $a=340,\phi=0.4$ and $a=320nm,\phi=0.2$ are provided for reference (grey dashed lines). The refractive index used for the chitin phase is 1.56.} 
    \label{fig:Figure4}
\end{figure}

When immersed in a medium with a refractive index greater than 1.30, single green scales appear predominantly red (Figure~\ref{fig:Figure4}A--B). HSM measurements across ten crystallites from each scale show a red shift of $\approx$90 nm in the mean peak reflectance wavelength between scales measured in air (mean peak wavelength = 526 nm) and scales measured in an immersion oil of refractive index 1.30 (mean peak wavelength = 616 nm; Figure~\ref{fig:Figure4}B--C). Increasing the refractive index further red-shifted the peak to mean peak wavelengths of 651 nm, 664 nm, and 678 nm for oil with refractive indices of 1.40, 1.45, and 1.50, respectively (Figure~\ref{fig:Figure4}B--C). Using FDTD simulations, we modelled the reflectance of three different idealised gyroid crystals of chitin in different refractive index environments (grey dashes and black line and points in Figure~\ref{fig:Figure4}C). The three models were gyroid crystals with lattice parameter equal to 320 nm, 330 nm, and 340 nm, with solid volume fractions of 0.2, 0.3, and 0.4, respectively. The model comprising the gyroid crystal with lattice parameter 330 nm, produced reflectance values that most closely matched our experimental data, but the simulated peak reflectance wavelengths are slightly (<10 nm) red-shifted in comparison with the measurement. All simulations however, supported the observed trend.

\section{Discussion}
The discrete and variably sized gyroid crystallites of \textit{E. opisena} offer an opportunity to gain insights into the formation of these remarkable structures \cite{WiltsScienceAdv2017}. In line with the findings from ref.\,\cite{WiltsScienceAdv2017}, we have shown that the crystallite areas are largest towards the tip ($\approx$40 $\mu$m$^2$) of the scale and smallest towards the base ($\approx$3 $\mu$m$^2$), using both SEM (Figure 3) and X-ray tomography (Figure S3). Using HSM, we have also shown that a positive correlation exists between the crystallite area and the intensity of reflected green wavelengths (Figure 3G). Full-wave optical modeling further supported these observations, but also hinted that, in addition to the crystallite area, the thickness of the crystallites plays a role (Figure~\ref{fig:Figure3}G). 

An analysis of X-ray tomography data showed that larger crystallites (in terms of area) were also thicker and both the area and thickness values obtained from the tomography data corresponded to our model parameters. By combining our tomography measurements with the modelling results, we show that it is possible to infer the thickness of each crystallite from their reflectance, and suggest that the larger crystallites at the scale tip are also thicker. Further investigations, such as a correlative tomography and spectroscopy study, would help to clarify this observation and enable more accurate predictions of crystallite sizes based on reflectance.

\subsection{Elucidating biophotonic nanostructure development using immersion experiments}
Our current understanding of biophotonic nanostructural development in butterflies is mostly based on studies using either fixed pupal wing tissue from discrete time points, or from adult wing scales \cite{ghiradella1974development,ghiradella1989structure}. Electron microscopy studies of fixed pupal wing tissues have offered insights into some of the more complex nanostructural formations such as the formation of lamella ridges and the growth of single gyroid photonic crystals \cite{ghiradella1974development, ghiradella1989structure, Saranathan2010}. More recently, studies using confocal microscopy have uncovered details of the cellular dynamics of scale formation including the role of F-actin in the formation of scale fingers and ridges \cite{Dinwiddie2014,lloyd2023actin}, in the spacing of ridges \cite{day2019sub}, and in the formation of the honeycomb lattice found in the upper lamina of some Papilionid butterfly species \cite{seah2023hierarchical}.

Many details of the developmental processes underlying nanostructure formation in butterflies remain elusive and can only be uncovered through continuous, spatio-temporally resolved \textit{in vivo} investigations of living scale cells. However, imaging nanostructural features \textit{in vivo} remains an unsolved challenge. Constrained by Abb{\'e}'s diffraction limit, the spatial resolution of light-microscopy is limited to hundreds of nanometers, a resolution too low to directly observe the development of photonic nanostructural features. In the last two decades, several super-resolution microscopy techniques utilising fluorescent probes have emerged, enabling \textit{in vivo} imaging with resolutions as low as 50 nm (for a review, see \cite{jing2021super}). While these techniques are promising, non-toxic fluorescent probes are scarce and expensive, and the strong illumination required by these techniques can easily disturb delicate subcellular processes, particularly when imaging must be performed over a span of several hours or days. Recently, a label-free imaging technique using optical speckle-correlation reflection phase microscopy achieved a resolution high enough to resolve some of the larger nanostructural changes, showing ridge and lamellae formation \textit{in vivo} \cite{McDougalKolle2021PNAS}, however the technique is not able to achieve the resolution needed to resolve the smaller gyroid nanostructures.

An alternative approach, suggested by this paper, is to take advantage of the fact that these nanostructures behave as photonic structures. As this paper has shown, it is possible to infer information about crystal size, and refractive index contrast from the optical signals that are created by photonic nanostructures. Importantly, our results suggest that HSM may be a convenient method for studying the small optical signals one would expect to occur during the formation of biological photonic crystals \textit{in vivo}. While MSP can provide similar spectral information, the HSM technique that we have applied in this study allows accurate measurements of spectra at the spatial resolution limit of optical microscopy while also efficiently providing spatial information (Figure~\ref{fig:Figure2}). For \textit{in vivo} imaging, this spatial information will be a necessity as keeping measurement consistency may be difficult during the inevitable movements caused by larval growth and HSM enables spatial registration of an image series.

Specifically, by immersing adult wing scales in refractive index-matching oils with similar refractive index to cytosol ($1.36 < n < 1.39$, \cite{liu2016cell}), we attempted to mimic the refractive index conditions that are likely to exist in a developing gyroid nanostructure. A reduction in the refractive index contrast between the chitin nanostructure and its surrounding medium causes a red-shift and a reduction of the reflectance (Figure 4). We additionally simulated this scenario using optical modelling on three different idealized gyroid nanostructures with varying lattice parameters and solid volume fractions. Our simulations of a gyroid nanostructure with lattice parameter equal to 330 nm and a solid volume fraction equal to 0.3, resulted in a near perfect match to our experimental data (Figure 4C). Lattice parameter and solid volume fraction are not easily obtained with high accuracy from the SEM images, but the best fitting values from our optical models correspond to those values reported previously from a related species \cite{schroder2011chiral, WiltsScienceAdv2017}.

Whether the red-shift also occurs in the developing butterfly remains to be seen, but these experimental simulations demonstrate the capacity of HSM to locally and accurately measure small reflectance signals. These results on mature wing scale nanostructures highlight that HSM has the appropriate spatial, temporal, and wavelength resolution to study the optical signals expected in developing butterfly scales.

\subsection{Hyperspectral microscopy as a tool to obtain spatially resolved reflectance}

The HSM technique implemented here uses a monochromatic camera and a liquid crystal tunable filter. These two devices can be easily added to most standard optical microscopes (Figure~\ref{fig:Figure1}). Tunable filters have been used previously in a variety of contexts, for example, to rapidly resolve variations in electroluminescence of light-emitting diodes \cite{edwards2023electroluminescence}, to obtain simultaneous low-resolution spectrophotometry of multiple stars \cite{slawson1999hyperspectral}, for fine arts diagnostics \cite{sapia2022hyperspectral}, in fluorescence microscopy \cite{zhang2023ultrathin}, and to analyze reflectance of butterfly wings \cite{medina2014detailed}. Their ease of use and relatively low cost makes them an attractive option for an \textit{in vivo} experimental set up.

Comparing HSM to MSP techniques, several advantages and limitations become apparent: HSM offers enhanced spatial resolution, allowing access to smaller measurement spots and providing statistical power that is difficult and laborious to obtain with MSP (Figure~\ref{fig:Figure3}). The superior spatial resolution brings about a trade-off with limited spectral range, reduced spectral resolution and slower acquisition rate. The spectral resolution is generally not a problem, as reflectance spectra of photonic structures are usually quite smooth. Particularly for UV-VIS-NIR spectral analysis and rapid data acquisition, wavelength range extension, increase in response speed and reduction of the sidebands (Figure~\ref{fig:Figure5}B) would be welcome. The wavelength range of HSM can be matched to the transmission limit of microscope optics by using a more traditional combination of a broadband (XBO) source, and a monochromator or a gradient bandpass filter, but this also brings about the necessity of focus control due to chromatic aberrations (\cite{blake2019compound,ilic2022simple}). Off-the-shelf multichannel LED sources based on dichroic mirrors, or multi-emitter LEDs with adequate spatial homogenisation are usable if the wavelength resolution is not of prime importance. Narrower spectra can be achieved by combining the LEDs \textit{via} a diffraction grating \cite{beluvsivc2016fast}. Compared to these solutions, the liquid crystal filter technology is attractive because it is lightweight, transportable and can be easily installed both in the illumination and the observation path.

In conclusion, the HSM technique described in this study enables the simultaneous recording of spectra across space in dynamic systems such as a developing butterfly, where temporal, spatial, and spectral information must be correlated. If the optical signal can be reliably measured \textit{in vivo} during wing scale development, hyperspectral microscopy may facilitate indirect measurements of structure formation and add new perspectives on processes involved in the growth of these intricate and intriguing natural optical elements.

\section{Methods}

\subsection{Samples}
\textit{Erora opisena} (Druce, 1912; Lepidoptera: Lycaenidae: Theclinae: Eumaeini) butterflies were caught in Chiapas, Mexico and purchased from The Bugmaniac (www.thebugmaniac.com). Individual wing scales were gently lifted off the wings using a cotton swab and placed onto a glass microscope slide or an aluminium stub covered with black aluminium foil (BKF12, T205, Thorlabs) or copper tape. Whole wings displayed in Figure~\ref{fig:Figure1} were imaged using a Leica M205 C digital microscope. Scanning electron microscopy images shown in Figure~\ref{fig:Figure1} were imaged using a FEI Verios XHR scanning electron microscope after wing scales were sputter coated with platinum. The electron beam acceleration voltage was set to 5\,kV and an in-lens secondary electron detector (Everhardt Thornley Detector) was used to image the scale at a working distance of approximately 5.0\,mm.

\subsection{Microspectrophotometry and hyperspectral microscopy}
A modified Zeiss Axioscope 5 optical microscope was used to perform both microspectrophotometry (MSP) and hyperspectral microscopy (HSM) measurements. Microscope objectives included Zeiss EC Epiplan-Apochromat objectives with a magnification of 50$\times$ (NA = 0.95), 20$\times$ (NA = 0.60), and 10$\times$ (NA = 0.30) for the measurements of Figures~\ref{fig:Figure2}--\ref{fig:Figure3} and a multi-immersion objective Zeiss 25$\times$ (NA = 0.80) LD LCI Plan-Apochromat for the measurements in oils presented in Figure~\ref{fig:Figure4}. 

For MSP, illumination was provided by a halogen light source (OSL2, Thorlabs Inc.) passed through a collimating lens (COP4-A, Thorlabs). Relative reflectance spectra were collected via a microscope sideport below the tube lens, with the light path consisting of a mirror, a focusing quartz lens and a 200 $\mu m$ quartz fibre (FC-UVIR200, Avantes) attached to a spectrometer (AvaSpec-ULS2048XL-EVO, Avantes). The measurement spot diameter of 4, 10, or 20 $\mu m$ using the 50$\times$, 20$\times$ and 10$\times$ objectives, respectively, was set to the centre of the image using a mirror in the object plane. An aluminium mirror (PF10-03-F01, Thorlabs) served as the reference.

\begin{figure}[ht]
    \centering
    \includegraphics[width=14cm]{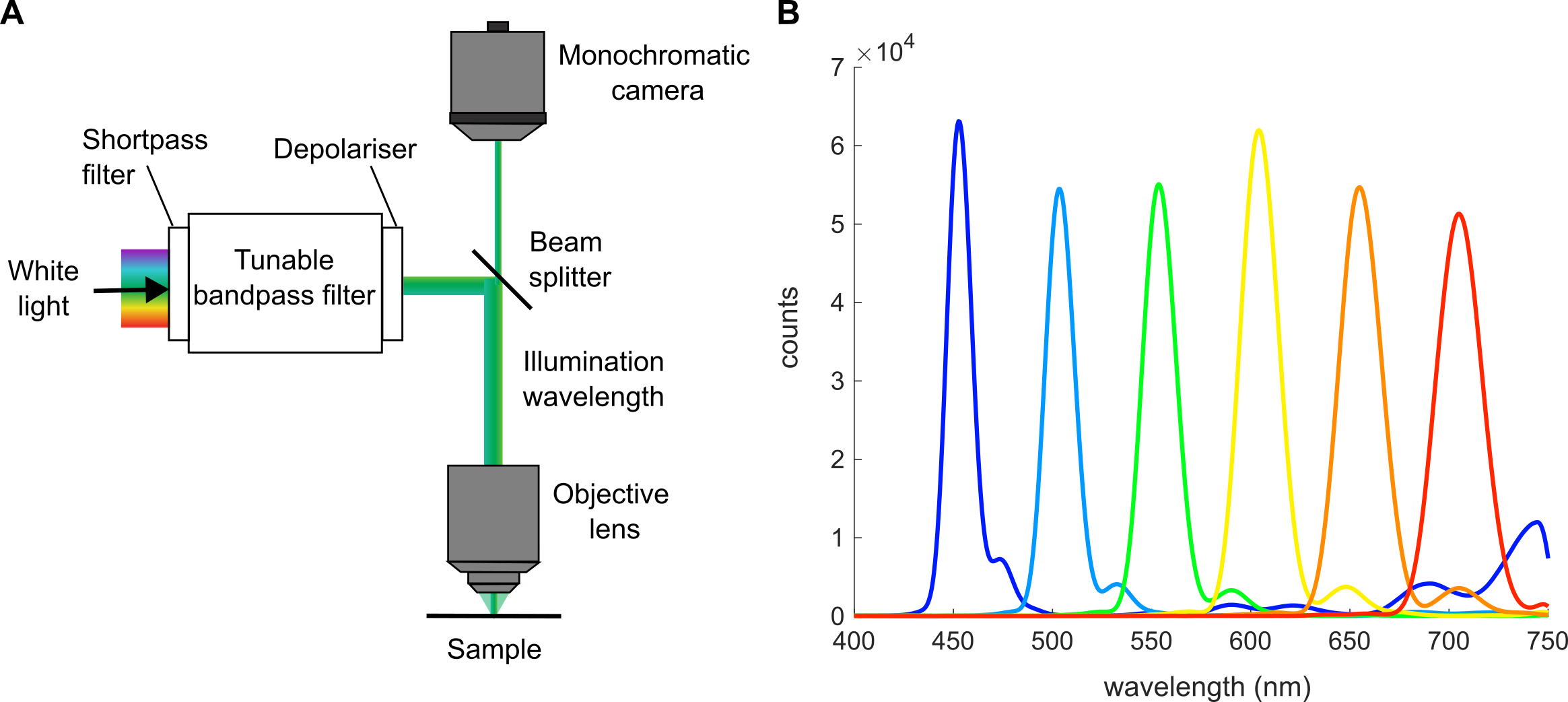}
    \caption{Hyperspectral microscopy set up and spectral output. (A) Schematic of the hyperspectral microscope consisting of a light source, a liquid crystal tunable bandpass filter, a monochromatic camera, and an optical microscope (see Methods for details). (B) Intensity of incident broadband light after filtering with the liquid crystal filter for centre wavelengths of 450--700 nm in 50 nm steps measured using a spectrometer.}
    \label{fig:Figure5}
\end{figure}

For HSM, illumination was provided by a Zeiss Axioscope 5 microscope collimated white LED source or a halogen light source (OSL2, Thorlabs) that was passed through a collimating lens (OSL2COL, Thorlabs) and a shortpass filter (FESH0750, Thorlabs). The collimated light was passed through a tunable liquid crystal bandpass filter (Kurios, Thorlabs Inc.), and a quartz depolariser to counteract any polarisation effect of the liquid crystal filter (DPU-25, Thorlabs; a schematic diagram is shown in Figure~\ref{fig:Figure5}A). The tunable bandpass filter was controlled over the serial interface by custom software written in MATLAB 2023a (The MathWorks Inc., USA). The bandwidth was set to deliver a restricted spectrum of light with a full-width at half-maximum (FWHM) of approximately 18\,nm (Figure~\ref{fig:Figure5}B). Samples were illuminated with wavelengths between 430--720 nm in 10 nm steps. For each illumination wavelength, an image was captured with a 20 MP monochromatic camera with pixel size 2.4 $\mu m$ (BFS-U3-200S6M-C, Teledyne FLIR) mounted to the camera port of the microscope. Because the illumination intensity (Figure~\ref{fig:Figure5}B) and the camera sensitivity was not equal across all wavelengths, a calibration run was initially conducted on a reference aluminium mirror (PF10-03-F01, Thorlabs) and the exposure times and gain values of the camera were adjusted such that the average pixel intensity of each image was approximately equal. This ensured that pixels were neither over nor underexposed for any one wavelength. Subsequently, the sample was measured using the calibrated exposure and gain values and the reflectance at each wavelength was calculated by averaging the pixels in the area of interest, subtracting the dark background from the sample data, and then dividing the sample data by the reference mirror to obtain relative reflectance. 

RGB images were taken on the same microscope using a 20 MP colour camera (The Imaging Source, DFK 38UX304) from different focal planes. Focus stacked images were then created using Helicon Focus Pro (Helicon Soft Ltd, Ukraine). All data was further analyzed using custom scripts written in MATLAB 2023a (The MathWorks Inc., USA).

\subsection{Correlation between reflectance and gyroid crystallite size}
To investigate the effect of crystallite size on reflectance we measured the average reflectance of regions of interest from scale base to tip and the reflectance of individual gyroid crystallites of differing sizes. A single scale was placed onto black foil attached to an aluminium stub and HSM was performed using the 50 $\times$ objective. Subsequently, the same scale was sputter-coated with a thin layer ($\approx$5--8 nm thick) of gold using a Cressington Sputter Coater 108 auto (120 s, 40 mA, background pressure 0.08 mbar) and imaged with a Zeiss Ultra Plus 55 scanning electron microscope (SEM). The electron beam acceleration voltage was set to 5\,kV and an in-lens secondary electron detector was used to image the scale at a working distance of approximately 5.5\,mm. SEM images were acquired at high magnifications across different sections of the scale and subsequently stitched together using Adobe Photoshop 2023. HSM and SEM data were then manually superposed. The regions of interest displayed in Figure~\ref{fig:Figure3}A--C each contain 160,000 camera pixels that were averaged to obtain reflectance spectra for each region. The regions of interest shown in Figure~\ref{fig:Figure3}E--F were defined from the SEM image using the Image Segmenter App in MATLAB 2023a.

To gain an understanding of how crystallite area (as viewed in the SEM image) relates to the thickness of the crystallite, we measured the areas and thicknesses of 25 crystallites from two X-ray tomograms of a green \textit{E. opisena} wing scale. Details of the X-ray tomography data collection can be found in \cite{WiltsScienceAdv2017}, and the tomography data sets are accessible at: \url{10.5061/dryad.w0vt4b912}. Briefly, an isolated wing scale was imaged using a Zeiss Xradia 810 Ultra X-ray microscope producing two tomography datasets with voxel sizes of 64 nm, one dataset of the base of the wing scale and a second of the tip of the wing scale. The datasets were visualized and stitched together in Dragonfly (v2022.1.0.1249, Object Research Systems, Montreal, Canada) and subsequent analyses were conducted on the stitched data. Area was measured manually on slices of the dataset using the free-hand drawing tool in Dragonfly (Figure S3), thickness was measured on two perpendicular slices using the measurement tool and the average of the two values was used as the crystallite thickness.

\subsection{Immersion oil experiments}
Single green wing scales were immersed in refractive index matching oils (Series A, Cargille Laboratories Inc.) with nominal refractive indices of 1.30, 1.40, 1.45, and 1.50 at 589.3\,nm at 25$^\circ$C. The mean reflectance of each scale immersed in the particular oil was obtained by averaging HSM measurements from ten randomly selected gyroid crystallites. Each measurement area was a square containing 100 camera pixels. 

\subsection{Finite-difference time-domain modelling}
Light scattering by the gyroid nanostructures was simulated with the three-dimensional finite-difference time-domain (FDTD) method, using Lumerical 8.29 (Lumerical Solutions, Vancouver, Canada). The gyroid nanostructures of \textit{E. opisena} were approximated by simulating voxelised single gyroid geometries in Houdini FX 19.5 (Side Effects Software Inc, Toronto, Canada) or via an idealized single gyroid network approximated by triply periodic minimal surface model from its level-set equation \cite{wohlgemuth2001triply}. Three gyroid geometries were simulated with lattice parameters of 320 nm, 330 nm, and 340 nm, and solid volume fractions of 0.2, 0.3, and 0.4, respectively. Gyroid geometries were set up in a rectangular simulation box with the two lateral directions, corresponding to [110] and [100], having periodic boundary conditions and the third (vertical) direction corresponding to the incident light direction was also a [110] symmetry direction having a perfectly matched layer (PML) boundary condition. The gyroid geometry that most closely matched our experimental data consisted of repeating unit cells with lattice parameter 330 nm and a solid volume fraction 0.3, resulting in the peak reflectance wavelength of 525 nm, approximately equal to that of our experimental reflectance measurements. %\sout{We assumed that} 
The gyroid consisted of cuticular chitin with a refractive index of 1.56 \cite{leertouwer2011refractive}, and we varied the refractive index of the surrounding medium between 1.00 and 1.55. Simulations of area and thickness dependence used PML boundary conditions on all boundaries of the simulation box that was adjusted in size to fully fit the gyroid geometry. Light, with wavelengths of 400-800\,nm, was incident in normal direction onto the structure along the [110] direction.

\section{Acknowledgements}
The authors acknowledge the facilities of Microscopy Australia at the Centre for Microscopy, Characterisation and Analysis, The University of Western Australia, a facility funded by the University, State and Commonwealth Governments. We thank Dr.\ Monte Masarei for assistance with image presentation, Dr.\ Alexandra Suvorova for assistance with SEM training and sample preparation, Dr.\ Volker Framenau for providing access to the Leica microscope, and Dr.\ Gregor Zickler for SEM support. We thank Dr.\ Neil Gale (Aberystwyth Butterfly House) for pointing out reference \cite{DryOutButterflyAber2024}. We thank Dr.\ Stephen Kelly (Zeiss), Dr.\ Benjamin Apeleo-Zubiri and Prof.\ Erdmann Spiecker (Erlangen) for the tomography data sets used in Fig.\ S4 in the supplementary material.

This work was supported by the Australian Research Council (ARC) through the Discovery Project DP200102593. This work was supported by a Research Grant from HFSP, Ref.-No: RGP0034/2021 (to B.D.W.). 

\printbibliography

\end{document}